 \documentclass[twocolumn,runningheads,natbib]{svjour2}

\def\xmm{{\sl XMM}-Newton}
\def\chan{{\sl Chandra}}
\def\psra{PSR~J0437--4715}

\def\fp{f_{\rm p}}
\def\ed{\dot{E}}
\def\lnon{L^{\rm nonth}}

\def\tpcc{T_{\rm pc}^{\rm core}}
\def\rpcc{R_{\rm pc}^{\rm core}}
\def\tpcr{T_{\rm pc}^{\rm rim}}
\def\rpcr{R_{\rm pc}^{\rm rim}}

\def\lbolpc{L_{\rm bol}^{\rm pc}}
\def\epc{\eta^{\rm pc}}
\def\enon{\eta^{\rm nonth}}
\def\lnon{L^{\rm nonth}}
\def\Log{{\rm Log}\,}
\newcommand{\gapr}{\raisebox{-.6ex}{\mbox{
$\stackrel{>}{\mbox{\scriptsize$\sim$}}\:$}}}
\newcommand{\lapr}{\raisebox{-.6ex}{\mbox{
$\stackrel{<}{\mbox{\scriptsize$\sim$}}\:$}}}
\newcommand{\vv}[1]{\mbox{\boldmath $#1$}}

\smartqed  
\usepackage{graphicx}
\journalname{Astrophysics and Space Science}
\begin{document}

\title{Studying Millisecond Pulsars in X-rays
}


\author{Vyacheslav E. Zavlin 
}


\institute{V.E. Zavlin \at
              Space Science Laboratory, NASA MSFC SD50, Huntsville, 
AL 35805, USA \\
              Tel.: +1-256-961-7463\\
              Fax: +1-256-961-7522\\
              \email{vyacheslav.zavlin@msfc.nasa.gov}           
}

\date{Received: date / Accepted: date}

\maketitle

\begin{abstract}
Millisecond pulsars represent an evolutionarily
distinct group among rotation-powered
pulsars. Outside the radio band, the soft X-ray range ($\sim 0.1$--10 keV)
is most suitable for studying
radiative mechanisms operating in these fascinating objects.
X-ray observations revealed 
diverse properties of emission from millisecond
pulsars. For the most of them, 
the bulk of radiation is of a thermal origin,
emitted from small spots (polar caps) 
on the neutron star surface heated by relativistic particles
produced in pulsar acceleration zones. On the other hand,
a few other very fast rotating pulsars exhibit almost pure nonthermal
emission generated, most probably, in pulsar magnetospheres. 
There are also examples of nonthermal emission 
detected from X-ray nebulae powered by millisecond pulsars, as well as
from pulsar winds shocked in binary systems with millisecond pulsars
as companions. These and other most important results obtained from
X-ray observations of millisecond pulsars are reviewed in this paper,
as well as results from the search for millisecond pulsations in
 X-ray flux of the radio-quite neutron star RX~J1856.5--3754.

\end{abstract}

\begin{figure}
\centering
\includegraphics[height=8cm,angle=0]{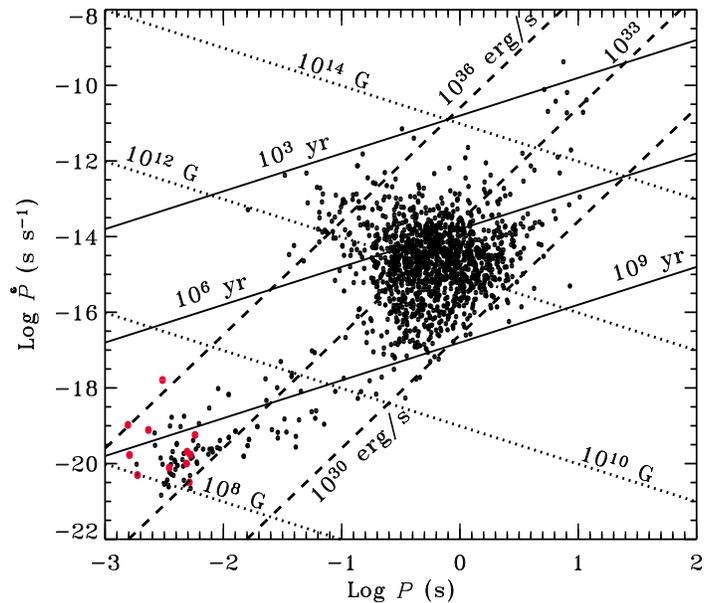}
\caption{$P$--$\dot{P}$ diagram for about 1,500 radio pulsars
(dots). 
Millisecond pulsars are located in the lower-left corner of the diagram.
The pulsars from Table~1 and 3 are shown with red dots.
Straight lines correspond to constant values of pulsar characteristic age
$\tau=10^3$, 10$^6$ and 10$^9$ yr, 
surface magnetic field $B_{\rm surf}=10^8$, 10$^{10}$,
10$^{12}$ and 10$^{14}$ G, and spin-down energy $\ed=10^{30}$,
10$^{33}$ and 10$^{36}$ erg s$^{-1}$.
}
\label{fig:1}   
\end{figure}

\begin{table}[t]
\begin{center}
\caption{Main parameters of eight MSPs}
\label{tab:1}
\begin{tabular}{llllll}
\tableheadseprule\noalign{\smallskip}
PSR & \,\,\,$\,\,\,\,P$ & \,\,\,$\,\,\,\,d$ &
\,\,\,$\,\,\,\,\,\,\,\,\,\,\,\,\tau$ & \,\,\,\,\,\,$\,\,\log\ed$    \\
 & \,\,\,(ms) &  \,\,\,(kpc) & \,\,\,\,\,\,\,\,\,(Gyr)  &
\,\,\,\,\,\,(erg\,s$^{-1}$)  \\[3pt]
\tableheadseprule\noalign{\smallskip}
B1937+21       & \,\,\,1.56 & \,\,\,3.57 &  \,\,\,\,\,\,\,\,\,\,\,\,0.24 & \,\,\,\,\,\,\,\,\,36.0     \\
B1957+20       & \,\,\,1.61 & \,\,\,2.49 &  \,\,\,\,\,\,\,\,\,\,\,\,2.24 & \,\,\,\,\,\,\,\,\,35.0    \\
J0218+4232     & \,\,\,2.32 & \,\,\,2.67 &  \,\,\,\,\,\,\,\,\,\,\,\,0.48 & \,\,\,\,\,\,\,\,\,35.4  \\
B1821--24      & \,\,\,3.05 & \,\,\,3.09 &  \,\,\,\,\,\,\,\,\,\,\,\,0.03 & \,\,\,\,\,\,\,\,\,36.3   \\
\noalign{\smallskip}\hline\noalign{\smallskip}
J0030+0451     & \,\,\,4.87 & \,\,\,0.32 &  \,\,\,\,\,\,\,\,\,\,\,\,7.71 & \,\,\,\,\,\,\,\,\,33.5   \\
J2124--3358    & \,\,\,4.93 & \,\,\,0.27 &  \,\,\,\,\,\,\,\,\,\,\,\,6.01 & \,\,\,\,\,\,\,\,\,33.6   \\
J1024--0719     & \,\,\,5.16 & \,\,\,0.39 &  \,\,\,$>27.25$ & \,\,\,$<32.9$   \\
J0437--4715    & \,\,\,5.76 & \,\,\,0.14 &  \,\,\,\,\,\,\,\,\,\,\,\,6.51 & \,\,\,\,\,\,\,\,\,33.5   \\
\tableheadseprule\noalign{\smallskip}
\end{tabular}
\end{center}
\vskip 0pt
\end{table}

\section{Introduction}
Millisecond pulsars
(MSPs) significantly differ in properties from other (ordinary) radio pulsars.
First of all, MSPs possess very short and stable spin periods, 
$P\lapr 50$ ms, with extremely small period derivatives, 
$\dot{P}\lapr 10^{-18}$ s s$^{-1}$.
These two main parameters separate MSPs from the majority
of other pulsars, as illustrated in the $P$--$\dot{P}$ 
diagram\footnote{
Based on the 
pulsar catalog 
provided by
the Australia Telescope National Facility 
(Manchester et al. 2005) 
and 
available at
{\tt http://www.atnf.csiro.au/research/pulsar}.}
shown in Figure~1. According to the conventional
pulsar magnetic-braking model, MSPs are very old neutron
stars, with characteristic ages $\tau=P/2\dot{P}\sim 0.1$--10 Gyr,
and low surface magnetic fields
$B_{\rm surf}\simeq 3.2\times 10^{19}\,(P\dot{P})^{1/2}\lapr 10^{10}$ G
(see Fig.~1). 
They were presumably spun up by angular momentum transfer
during a mass accretion phase in binary
systems\footnote{It is why MSPs are often called ``recycled''
pulsars.}
(e.g., Alpar et al. 1982), 
and their low values of $B_{\rm surf}$ could be explained
by the Ohmic and/or accretion-induced decay of magnetic field
(see Cumming 2005 and references therein).
Discoveries of several accretion-driven binary MSPs, 
first of all the famous pulsar
SAX J1808.4--3658 with $P\simeq 2.5$ ms, 
support this hypothesis
(see, e.g., Wijnands 2004
for a review on accreting MSPs).

Since the discovery of the first fast
rotating pulsar B1937+21 by Backer et al. (1982),
MSPs have been extensively searched for and studied in radio domain.
Currently, about 130 MSPs are known (Manchester et al. 2005).
Outside the radio band, as
MSPs are intrinsically faint at optical
wavelengths and most of them ($\sim 80$\%) 
reside in binary systems with optically brighter
white dwarf companions, the soft X-ray energy range 
($\sim 0.1$--10 keV) is the main source of
information on these pulsars.
The detection of pulsed X-ray emission from the brightest (and
nearest) MSP J0437--4715 with $ROSAT$ (Becker \& Tr\"umper 1993)
initiated a series of dedicated X-ray observations of these
intriguing objects in 90's
with this satellite and also $ASCA$, Beppo$SAX$ and $RXTE$.
Later on this observational ``relay'' has been continued with
\chan\ and \xmm.
So far, firm X-ray detections have been reported
for about three dozens of isolated (solitary, or non-accreting
if in binaries) MSPs.
The majority of these objects are located in 
globular clusters (mostly  
47~Tuc --- see
Bogdanov et al. 2006).
This paper mainly concentrates on eight MSPs, listed in Table~1, with
available detailed
information on properties of detected
X-ray emission.
Besides $P$ and $\tau$,
Table~1 gives estimates on  the distances to these objects, $d$ 
(inferred from either pulsar parallaxes or dispersion measures
and the NE2001 galactic electron density model
by Cordes \& Lazio 2003)\footnote{
For the MSPs of the second group in Table~1 (with $d<1$ kpc),
the distance estimates obtained from the pulsar parallaxes and
those inferred from their dispersion measures agree well with
each other (see Hotan, Bails \& Ord 2006 and Lommen et al. 2006).},
and on their spin-down energies, $\ed=4\pi^2 I P^{-3}\dot{P}$
(assuming a standard neutron star moment of inertia, $I=10^{45}$ g cm$^2$). 
Note that for PSR J1024--0719 the intrinsic period derivative
is not well determined, 
$\dot{P}<3\times 10^{-21}$ s s$^{-1}$ (Hotan, Bails \& Ord 2006),
that results in the lower and upper limits on $\tau$ and $\ed$,
respectively.

Generally, X-ray emission from radio pulsars consists of two
different components, thermal and nonthermal, generated 
on the neutron star surface or in its vicinity.
The nonthermal component is usually
described by a power-law (PL) spectral model and attributed to
radiation produced by synchrotron and/or inverse Compton
processes in the pulsar magnetosphere, whereas the thermal
emission can originate from either the whole surface of a cooling 
neutron star
or small hot spots around the magnetic poles (polar caps; PCs)
on the star surface, or both.
As predicted by virtually all pulsar models, these PCs can be heated
up to X-ray temperatures ($\sim 1$ MK) by relativistic
particles generated in pulsar acceleration zones.
A conventional assumption about the PC radius is that it is close
to the radius within which open magnetic field lines originate from
the pulsar  surface,
$\sim [2\pi R_{\rm NS}^3/cP]^{1/2}\simeq 2\,[P/5\,{\rm ms}]^{-1/2}$ km 
(for a neutron star radius $R_{\rm NS}=10$ km). 
In case of MSPs,
the entire surface at a neutron star age of $\sim 1$ Gyr is too cold,
$\lapr 0.1$ MK, to
be detectable in X-rays (although it may be seen in the $UV/FUV$ band ---
see Sec.~6).
Therefore, only nonthermal (magnetospheric) and/or
thermal PC radiation is expected to be observed in X-rays
from these objects. 
In addition to the radiation produced
by MSPs themselves, nonthermal emission from pulsar-wind
nebulae (PWNe)
associated with MSPs moving at supersonic
velocities ($\gapr 100$ km s$^{-1}$) through interstellar medium
may be detected. Another
source of nonthermal X-ray radiation generated in binary systems can be 
an intrabinary shock formed where the pulsar wind and
matter from the stellar component collide
(Arons \& Tavani 1993),
although such a component would be hardly separated
from radiation produced by the pulsar itself
(unless properties of the nonthermal emission varies with orbital phase).

The MSPs in Table~1 belong to two distinct groups:
those which emit almost pure nonthermal radiation
and those with a predominantly thermal PC component. 
The next two Sections provides details on the X-ray properties
of these objects. Section~4 is devoted to
three X-ray emitting MSPs whose properties remain uncertain.
X-ray PWNe powered by MSPs are briefly discussed in Section~5.
A summary is given in Section~6.

\begin{figure}
\centering
\includegraphics[height=7cm,angle=0]{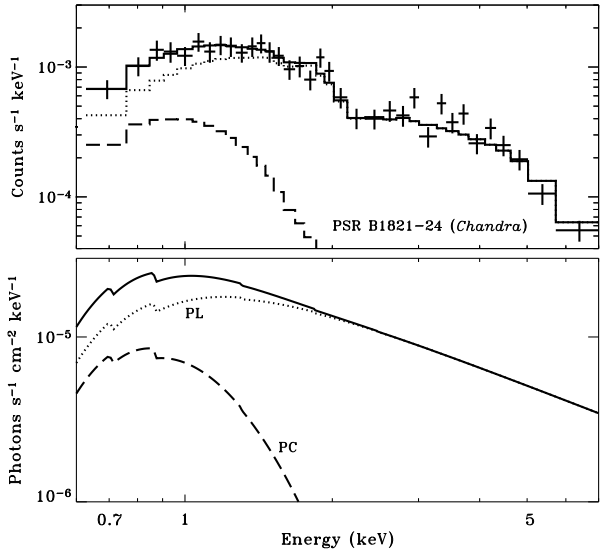}
\caption{X-ray spectrum of PSR B1821--24 as detected with the
\chan\ ACIS-S instrument
(crosses)
fitted with a PL model of $\Gamma=1.2$
(dotted curves) plus a possible PC component (dashes).
See also Becker et al. (2003).
}
\label{fig:2}
\end{figure}

\begin{figure}
\centering
\includegraphics[height=7cm,angle=0]{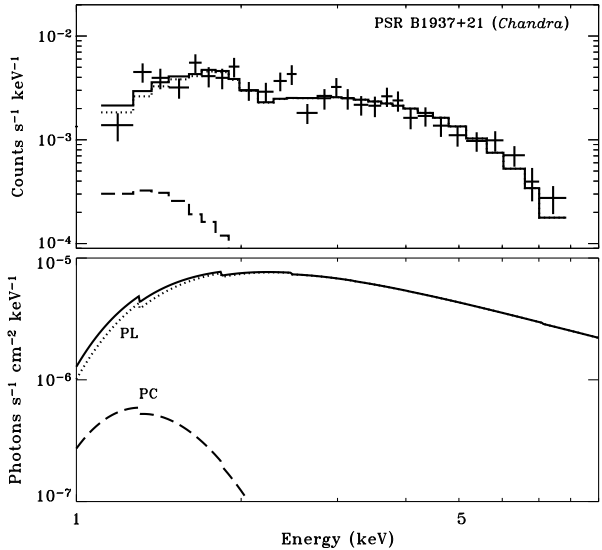}
\caption{Same as in Fig.~2 for PSR B1937+21 and a PL model of $\Gamma=1.2$
(see also Kuiper et al. 2006).
}
\label{fig:3}
\end{figure}

\begin{figure}
\centering
\includegraphics[height=7cm,angle=0]{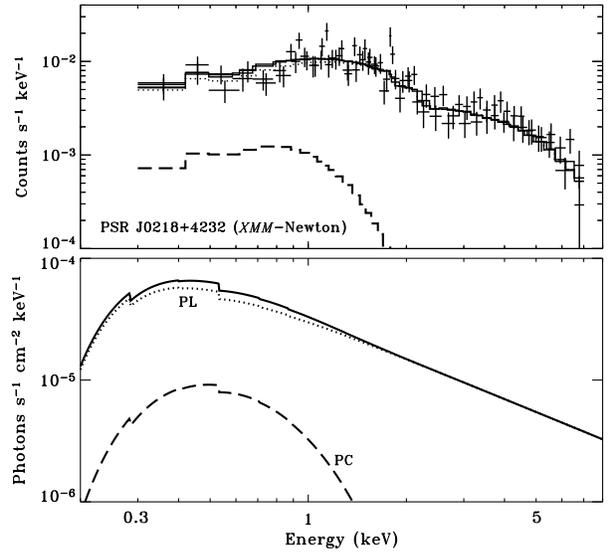}
\caption{Same as in Fig.~2 for PSR J0218+4232
(as detected with the \xmm\ EPIC-MOS instruments)
and a PL model of $\Gamma=1.1$ (see also Webb, Olive \& Barret 2004).
}
\label{fig:4}
\end{figure}

\begin{figure}
\centering
\includegraphics[height=7cm,angle=0]{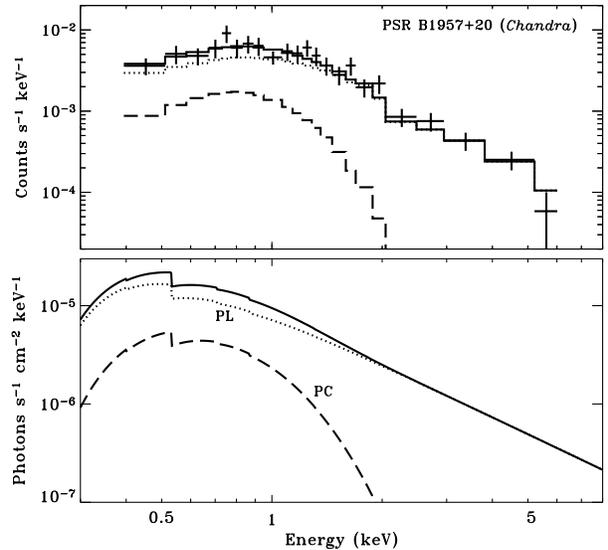}
\caption{Same as in Fig.~2 for PSR B1957+20 and a PL model of $\Gamma=1.9$.
(Note that the detected flux can contain a contribution
from a possible intrabinary shock --- see Sec.~2 and
Stappers et al. 2003 for more details.)
}
\label{fig:5}
\end{figure}


\section{Nonthermally emitting MSPs}
Of the MSPs listed in Table~1, these are 
PSRs B1937+21, B1957+20, J0218+4232, and B1821--24.
They are characterized by 
large values of spin-down energy,
$\ed>10^{35}$ erg s$^{-1}$, and shorter spin periods,
$P<3.1$ ms. PSR J0218+4232 is also one of few $\gamma$-pulsars known
(Kuiper et al. 2000).
X-ray observations of these MSPs revealed their nonthermal spectra,
albeit with different photon indices $\Gamma$. 
Figures~2--5 show the spectra detected with \chan\ and/or
\xmm\
from these MSPs.
Although the spectral data on these pulsars do not formally
require any other component (in addition to a PL),
one cannot exclude that there is also a contribution
of thermal PC emission in the detected X-ray fluxes,
as indicated in Figures~2--5.  
For PSRs B1821--24, J0218+4232, and B1937+21 
the measured photon indices are $\Gamma\simeq 1.1$--$1.2$. 
The spectrum of  
PSR B1957+20, was found
to be much steeper, with $\Gamma\simeq 1.9$
(although the X-ray radiation of this pulsar
can consist of two component --- see below).
Note that
these estimates on $\Gamma$ are derived for the phase-integrated fluxes, 
whereas the observational
data on PSRs J0218+4232 
and B1937+21 indicate that the spectral
slope may change with pulsar rotational phase (see Webb, Olive \&
Barret 2004 and
Nicastro et al. 2004).
However, much more sensitive observations are required to confirm
this effect. 
Regarding the X-ray spectrum of PSR B1821--24 (located in
the globular cluster M28), Becker et al. (2003)
speculated that there is a marginal evidence for an emission line
around 3.3 keV (see also Fig.~2). 
If this feature is real\footnote{
There is
no indication of a feature at this
photon energy in the spectra of PSRs B1957+20 and B1937+21
detected with the same \chan\ ACIS-S instrument (Figs. 3 and 5).}, 
it could be
interpreted as cyclotron emission from an optically thin corona 
above the pulsar provided its magnetic field is strongly different
from a centered dipole.
The estimated nonthermal (isotropic) luminosities\footnote{
X-ray luminosities of the pulsars discussed in Sections~2 and 3
are derived for the distances given in Table~1.}  
of these four objects in the 0.2$-$10 keV range, 
$\lnon$, and the corresponding ``nonthermal''
efficiencies, $\enon=\lnon/\ed$, are given in Table~2.
Note that a fraction of the X-ray emission detected
from the eclipsing 
pulsar B1957+20,
which is thought to ablate its companion
in a close binary system with a 9.2-hr orbital period,
may be due to an intrabinary shock between the pulsar wind
and that of the companion star. The \chan\ data on this MSP
showed an apparent (at a 99\% confidence level)
modulation of the X-ray emission detected at the
pulsar's position with orbital phase
(Stappers et al. 2003), with lowest and highest fluxes
during and immediately after eclipse (respectively).
Assuming that this flux modulation is genuine,
Stappers et al. (2003)
obtained a 50\% estimate on the
contribution of X-ray emission from the intrabinary shock in
the total flux detected from the pulsar. This fraction
is accounted for in deriving the nonthermal luminosity and
efficiency for PSR B1957+20 given in Table~2.

\begin{figure}
\centering
\includegraphics[height=9cm,angle=0]{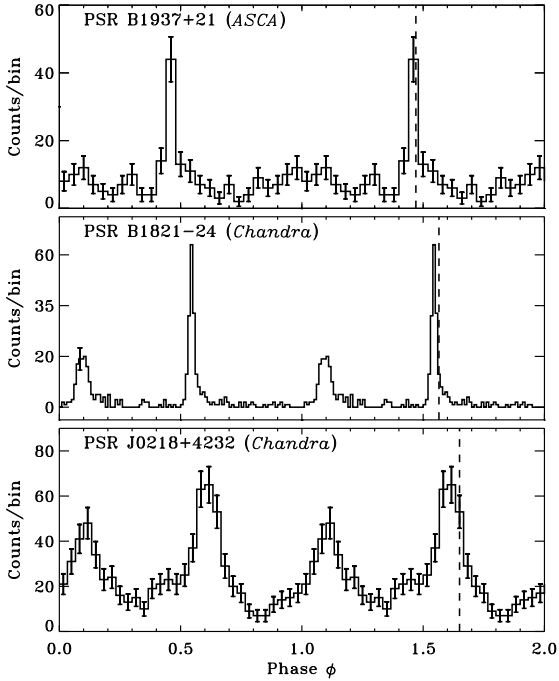}
\caption{
Pulsed profiles of three nonthermally emitting MSPs.
Vertical dashed
lines indicate phases of main radio pulses.
See Takahashi et al. (2000),
Rutledge et al. (2004), and Kuiper et al. (2002) for more details on
timing analysis of X-rays from PSRs B1937+21, B1821--24, and
J0218+4232 (respectively).
}
\label{fig:6}
\end{figure}

Another prominent feature of the nonthermal emission from three
of these MSPs is the shape of the X-ray pulsed profiles,
with strong and narrow main pulses and large pulsed fractions $\fp$
ranging from about 65\% for PSR B0218--4232 up to nearly 100\% for
PSRs B1937+21 and B1821--24. Figure~6 presents the
pulsars' light curves. 
Nonthermal pulsed emission from these MSPs
was also detected up to about 20 keV with $RXTE$
(Rots et al. 1998; Kuiper, Hermsen \& Stappers 2004; Cusumano et al. 2003).
The main X-ray and radio peaks of these pulsars
were found to be nearly aligned in phase
(based on both \chan\ and $RXTE$ data) 
\footnote{Although Takahashi et al. (2001) found from $ASCA$
data that the main
X-ray and radio pulses of PSR B1937+21 are separated by
$\Delta\phi\simeq 0.5$ in phase.}.
This suggests that nonthermal photons emitted from
these MSPs in the radio and X-ray bands  
are generated in the same (or close) zones,
although it still remains unclear where those zones are located:
closer to the neutron star surface (as suggested by the polar-cap model of
Harding, Usov \& Muslimov 2005) or near the pulsar light cylinder
(according to the outer-gap model --- e.g., Cheng, Ho \&
Ruderman 1986, Romani 1996).
Pulsations of the X-ray flux emitted by  
PSR B1957+20 have to be detected yet\footnote{No results from the
\xmm\ observation of this MSP conducted
in October 2004 have been reported yet.}.

\section{Thermally emitting MSPs}
This is the second group of objects in Table~1
which consists of PSRs J0030+0451, J2124--3358, J1024--0719,
and J0437--4715. These MSPs are less energetic in terms of $\ed$, have
longer spin periods and are much closer to Earth.
X-ray spectra of the two brightest members of
this group, PSRs J0030+0451 and J0437--4715
cannot be fitted with a single PL model, whereas for the other two
pulsars such a fit yields too large photon indices, $\Gamma\simeq 3$--4,
and the obtained estimates on hydrogen column density, 
$n_{\rm H}\simeq (1$--$2)\times 10^{21}$ cm$^{-2}$, towards these
objects greatly exceed those inferred from
independent measurements
(Zavlin 2006). A broken-PL model, suggested by Becker \&
Aschenbach (2002) to interpret the spectra of PSRs J0030+0451 and J0437--4715,
results in unrealistic estimates on $n_{\rm H}$. 
A model involving a PL and a simple one-temperature thermal component
also faces the same problem.

Analyzing $ROSAT$ data on \psra, Zavlin \& Pavlov (1998)  
suggested a thermal model implying a nonuniform temperature
distribution over PCs. These authors discussed that
relativistic particles bombarding magnetic poles of an MSP
could heat a region larger
than the conventional PC size (see Sec.~1) because the low magnetic field
of the pulsar does not prevent the released
heat from propagating along the neutron star surface.
The applied model assumes two identical PCs on magnetic poles 
covered with a weakly magnetized hydrogen atmosphere 
(Zavlin, Pavlov \& Shibanov 1996).
It also takes into account the GR effects
(redshift and bending of photon trajectories near the neutron star surface).
In this model the thermal radiation depends on PC temperature and radius, 
the neutron star
mass-to-radius ratio, which determines the GR effects, and the star geometry
(the viewing and magnetic angles, $\zeta$ and $\alpha$, respectively).
The sketch shown in Figure~7 illustrates the neutron star
geometry and the effect of light bending (see also, e.g., Zavlin, Shibanov
\& Pavlov 1995
for more details on the GR effects on pulsar PC emission).
The nonuniform temperature was approximated with a step-like function,
referred as PC ``core'' and ``rim''.
Observations of \psra\ with \chan\ (Zavlin et al. 2002) and
\xmm\ (Zavlin 2006) showed that such a model, supplemented with a PL component
of $\Gamma\simeq 2.0$, fits well the spectrum of \psra\ up to 10 keV and
yields reasonable pulsar parameters as well as $n_{\rm H}$.
The thermal model provides the bulk of the pulsar's X-ray flux, whereas
the PL component prevails only at photon energies
$E\gapr 3$ keV. A similar (nonuniform PCs plus PL) model works also well
on the spectral data of PSR J0030+0451. Figures~8 and 9 demonstrate
this. The inferred parameters of the thermal components are:
$\tpcc\simeq 1.4$ and 2.1 MK, $\tpcr\simeq 0.5$ and 0.8 MK,
$\rpcc\simeq 0.4$ and 0.1 km, $\rpcr\simeq 2.6$ and 1.4 km,
for PSRs J0437--4715 and J0030+0451, respectively (in the latter case
the PL index was fixed at $\Gamma=1.5$ because of a poorer data statistics
at higher energies). For PSRs J2124--3358
and J1024--0719 the same model yields PC parameters close to those
mentioned above, although low quality of the observational data
on these two objects at $E>3$ keV allows one to put only upper limits on
intensity of the nonthermal component (Zavlin 2006). The bolometric
luminosities of one PC, $\lbolpc$, and ``PC efficiencies'',
$\epc=\lbolpc/\ed$, can be found in Table~2,
together with the estimates on $\lnon$ and $\enon$ for the nonthermal
component (or corresponding 1$\sigma$ upper limits).
Upper limits on the luminosity of a possible thermal PC component in
the X-ray fluxes of the four nonthermally emitting pulsars
(Sec.~2) are also given in Table~2. 
Note that a more detailed study of possible thermal emission from these
four pulsars is hampered by large distances (a few kpc) and, hence, strong
interstellar absorption ($n_{\rm H}\sim [2$-$4]\times 10^{21}$ cm$^{-2}$)
towards these objects.

\begin{figure}
\centering
\includegraphics[height=8cm,angle=0]{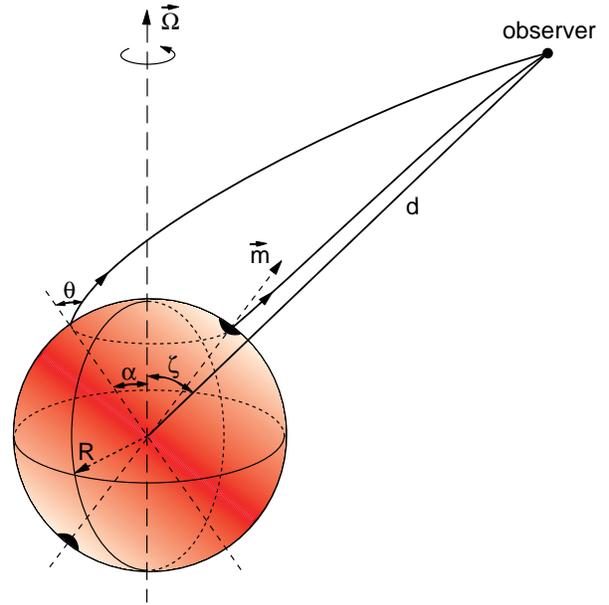}
\caption{Sketch illustrating bending of photon trajectories in a strong
gravitational field near the surface of a neutron star
with rotational
and magnetic axes $\vv{\Omega}$ and $\vv{m}$, and viewing and
magnetic angles $\zeta$ and $\alpha$ (respectively).
}
\label{fig:7}
\vskip -0pt
\end{figure}

\begin{figure}
\centering
\includegraphics[height=7cm,angle=0]{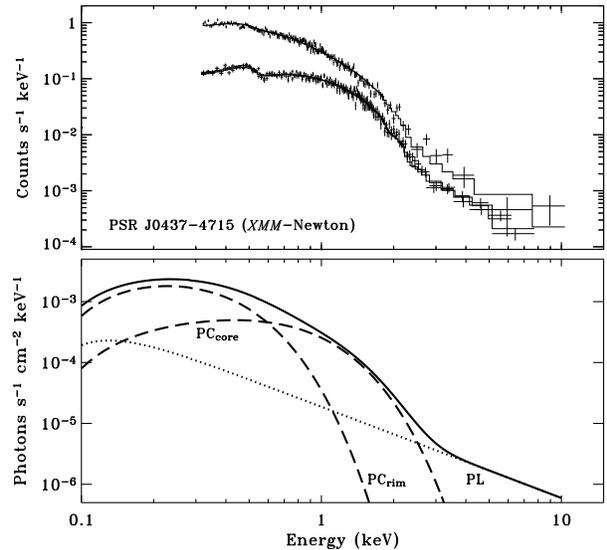}
\caption{X-ray spectra of \psra\ as detected with the \xmm\
EPIC-pn and MOS instruments (crosses).
The solid curves show a best fitting model, PL (dots) plus
a two-temperature  PC component (dashes),
``core'' and ``rim'' (see Sec.~3).
}
\label{fig:8}
\vskip -0pt
\end{figure}

\begin{figure}
\centering
\includegraphics[height=7cm,angle=0]{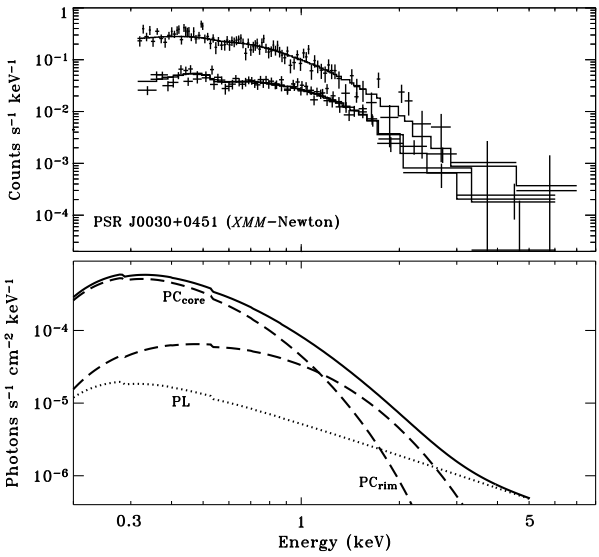}
\caption{Same as in Fig.~8 for PSR J0030+0451.
}
\label{fig:9}
\vskip -0pt
\end{figure}

\begin{figure}
\centering
\includegraphics[height=12cm,angle=0]{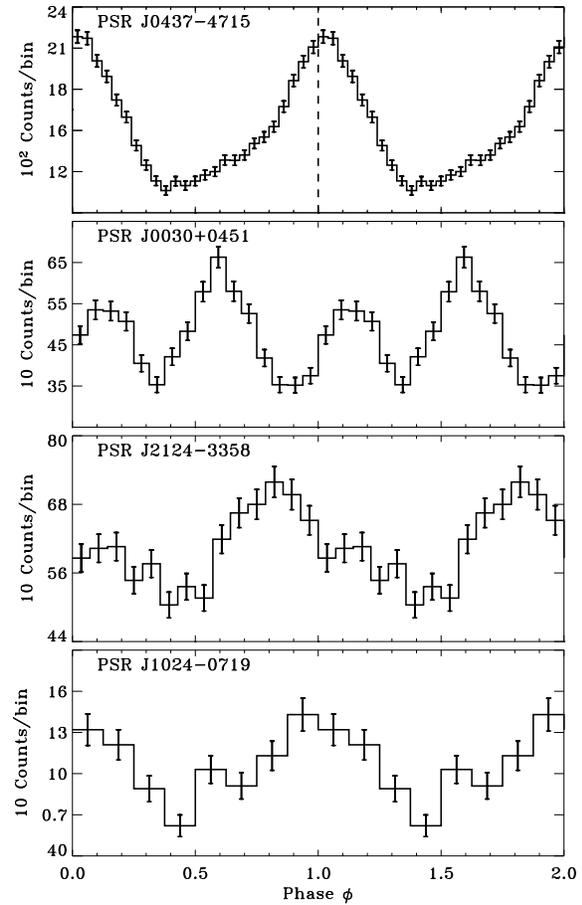}
\caption{
Pulsed profiles of thermally emitting MSPs
obtained from \xmm\ observations
(Zavlin 2006).
The vertical dashed
line in the upper panel indicates phase of a radio pulse of \psra\ 
(see Sce.~3 for details).
}
\label{fig:10}
\vskip -0pt
\end{figure}

As alternative to the PC-plus-PL model of the X-ray emission
from \psra, Bogdanov, Grindlay \& Rybicki (2006) suggested a purely
thermal interpretation in which the harder X-ray spectral
tail (at $E\gapr 3$ keV)
is a result of the inverse Compton scattering of soft thermal PC photons
by energetic electrons/positrons in an optically thin thermal layer
of a temperature $kT_{\rm e}\sim 150$ keV located presumably in the 
pulsar's magnetosphere. However, as
existence of such a thermal layer in a pulsar magnetosphere
does not seem to be physically supported, 
this Comptonization interpretation remains rather questionable.

\begin{table*}[t]
\begin{center}
\caption{X-ray luminosities of eight MSPs}
\label{tab:2}
\begin{tabular}{lllll}
\tableheadseprule\noalign{\smallskip}
PSR & \,\,\,\,\,\,\,$\Log\lnon$ & \,\,\,\,\,$\Log\lbolpc$ & \,\,\,\,\,\,\,\,\,$\Log\enon$ & \,\,\,\,\,\,\,\,\,$\Log\epc$  \\
 &  \,\,\,\,\,\,\,\,(erg\,s$^{-1}$) & \,\,\,\,\,(erg\,s$^{-1}$) & & \\[3pt]
\tableheadseprule\noalign{\smallskip}
B1937+21       &  \,\,\,\,\,\,\,\,\,\,\,32.8 & \,\,\,\,\,$<31.5$ & \,\,\,\,\,\,\,\,\,\,\,\,\,\,-3.2 &  \,\,\,\,\,$<-4.5$  \\
B1957+20       & \,\,\,\,\,\,\,\,\,\,\,31.5  & \,\,\,\,\,$<30.4$  & \,\,\,\,\,\,\,\,\,\,\,\,\,\,-3.5 & \,\,\,\,\,$<-4.6$  \\
J0218+4232     &  \,\,\,\,\,\,\,\,\,\,\,32.6 & \,\,\,\,\,$<31.0$ & \,\,\,\,\,\,\,\,\,\,\,\,\,\,-2.8 & \,\,\,\,\,$<-4.4$ \\
B1821--24      &  \,\,\,\,\,\,\,\,\,\,\,32.7 & \,\,\,\,\,$<32.0$ & \,\,\,\,\,\,\,\,\,\,\,\,\,\,-3.6 & \,\,\,\,\,$<-4.3$ \\
\noalign{\smallskip}\hline\noalign{\smallskip}
J0030+0451     & \,\,\,\,\,\,\,\,\,\,\,29.8 & \,\,\,\,\,\,\,\,\,\,\,30.3 & \,\,\,\,\,\,\,\,\,\,\,\,\,\,-3.7 & \,\,\,\,\,\,\,\,\,\,\,\,\,\,-3.2 \\
J2124--3358    & \,\,\,\,\,$<29.0$ & \,\,\,\,\,\,\,\,\,\,\,30.0 & \,\,\,\,\,$<-4.6$ & \,\,\,\,\,\,\,\,\,\,\,\,\,\,-3.6 \\
J1024--0719     & \,\,\,\,\,$<28.9$  & \,\,\,\,\,\,\,\,\,\,\,29.3 & \,\,\,\,\,\,\,\,\,\,\,\,\,\,\,...... & \,\,\,\,\,$>-3.6$  \\
J0437--4715    &  \,\,\,\,\,\,\,\,\,\,\,29.7 & \,\,\,\,\,\,\,\,\,\,\,30.2 & \,\,\,\,\,\,\,\,\,\,\,\,\,\,-3.8 & \,\,\,\,\,\,\,\,\,\,\,\,\,\,-3.3 \\
\tableheadseprule\noalign{\smallskip}
\end{tabular}
\end{center}
\end{table*}

X-ray emission from all these four thermally emitting MSPs is pulsed,
with pulsed fraction ranging from about 35\% up to 50\% 
(Becker \& Aschenbach 2002; Zavlin 2006). 
The pulsed profiles of PSRs J0437--4715, J2124--3358, and
J1024--0719 are rather similar in shape, with single 
broad pulses,  
whereas the light curve of PSR J0030 +0451 exhibits two
pulses per period separated by $\Delta\phi\simeq 0.5$ in phase. 
This indicates that the 
geometry of PSR J0030+0451 
(the angles $\zeta$ and $\alpha$) 
is different from those of the three others.
For example, in the framework of the conventional pulsar model
with the magnetic dipole at the neutron star center,
PSR J0030+0451 can be a nearly orthogonal rotator
(i.e., $\zeta\simeq\alpha\simeq 90^\circ$)
with two pulses in its light curve being due to contributions
from two PCs seen during the pulsar's rotation. For the other MSPs,
the bulk of the detected X-ray fluxes is expected to 
come mostly from one PC.
However, the X-ray pulses of PSRs J0437--4715, J1024--0719, and J2124--3358
are clearly asymmetric, with a longer rise and a faster decay
for the former two MSPs and with the opposite behavior for the latter pulsar.
None of these shapes can be explained
by a simple axisymmetric temperature distribution.
A feasible interpretation is that the observed asymmetry in the
pulsed profiles of these three MSPs is caused by contribution
of the nonthermal component whose peak is shifted in phase with respect
to the pulse of the thermal emission. Results of the energy-resolved
timing and phase-resolved spectroscopy on the \xmm\ data on
\psra\ support this explanation (Zavlin 2006).
For PSR J2124--3358, there may be an alternative interpretation
of the shape of the pulsar's light curve: the steeper rise and longer trail
could be caused by relativistic
effects (in particular, the Doppler boost) in fast rotating
pulsars 
(Braje, Romani \& Rauch 2000), although it then should be 
understood why these effects are not seen in the pulse profiles
of the other three MSPs whose the spin periods close
to that of PSR J2124--3358 (Table~1).
Another way to attempt to explain the observed light curves is to involve
a model with a decentered magnetic dipole
(or another magnetic field configuration ---
see the presentation by S. Zane at this conference), 
what has not been done yet for interpreting pulsed X-ray emission
from MSPs.
In any case, to produce a pulsed fraction at
a level of $\fp\gapr 30\%$,
thermal emission has to be 
anisotropic, as predicted by neutron star atmosphere models,
to counteract the effect of 
light bending near the star surface on pulsations of the thermal flux 
(Zavlin, Shibanov \& Pavlov 1995). 
So far, the difference in phases between radio
and X-ray pulses was determined only for \psra\ based on \chan\ data
(Zavlin et al. 2002; see also Fig.~10). 
As the bulk of the X-ray flux from this pulsar is
of the thermal PC origin, the small difference in phases of
the radio and X-ray peaks suggests
that the pulsar's radio emission is generated close to
the neutron star surface, unless the effects of field-line sweepback 
and/or aberration cancel the travel-time difference.

\begin{table*}[t]
\begin{center}
\caption{Three MSPs with estimates on X-ray flux}
\label{tab:3}
\begin{tabular}{lllllll}
\tableheadseprule\noalign{\smallskip}
PSR & \,\,\,\,$\,\,\,\,P$ & $\,\,\,\,\,\,\,\,d$ &
$\,\,\,\,\,\,\,\,\,\,\,\,\tau$ & 
\,\,\,$\,\,\log\ed$ & \,\,\,\,$\log{L_{\rm X}}$ &
\,\,\,\,$\log{\eta_{\rm X}}$   \\
 & \,\,\,\,(ms) &  \,\,\,\,(kpc) & \,\,\,\,\,\,(Gyr)  &
\,\,\,(erg\,s$^{-1}$) & \,\,\,(erg\,s$^{-1}$) & \\[3pt]
\tableheadseprule\noalign{\smallskip}
J0034--0534   & \,\,\,\,1.88 & \,\,\,\,0.54 &  \,\,\,\,\,\,\,\,\,6.0 & 
\,\,\,\,\,\,34.5 & 
\,\,\,\,\,\,29.7 & \,\,\,\,-4.8    \\
J0751+1807    & \,\,\,\,3.48 & \,\,\,\,0.63 &  \,\,\,\,\,\,\,\,\,7.7 & 
\,\,\,\,\,\,33.8 &  
\,\,\,\,\,\,30.3 & \,\,\,\,-3.5   \\
J1012+5307  & \,\,\,\,5.26 & \,\,\,\,0.41 &  \,\,\,\,\,\,\,\,\,8.6 & 
\,\,\,\,\,\,33.4 &
\,\,\,\,\,\,30.4 & \,\,\,\,-3.0 \\
\tableheadseprule\noalign{\smallskip}
\end{tabular}
\end{center}
\vskip 0pt
\end{table*}

\section{More X-ray emitting MSPs}
Three pulsars, J0034--0534, J0751+1807, 
and J1012+5307, are examples of MSPs firmly detected in X-rays but
whose radiative properties remain
uncertain. 
X-ray observations of these objects were not 
long enough to provide data suitable for 
discriminating among various
spectral models. Those data could be equally
well fitted with a PL model of $\Gamma\sim 1.7$
or a thermal (blackbody) model of a temperature
of $\sim 3$ MK (see Webb et al. 2004 and Zavlin 2006).
Estimates on the total flux are the only 
rather reliable X-ray characteristics available 
for these objects.
Table~3 presents the main pulsar parameters\footnote{
The distance to PSR J0751+1807 is derived from the pulsar's
parallax (Nice et al. 2005), whereas the other two estimates
are obtained from the pulsars' dispersion measures and the NE2001 model.}
together with the estimated X-ray luminosities in the
0.2--10 keV range, $L_{\rm X}$, and corresponding efficiencies,
$\eta_{\rm X}=L_{\rm X}/\ed$. 
In terms of the spin-down power $\ed$, PSRs J0751+1807
and J1012+5307 join the four
pulsars exhibiting thermal PC emission (Sec.~3),
while PSR J0034--0534 seems to be more energetic and closer
to the first group of the nonthermally emitting MSPs (Sec.~2).
On the other hand, PSR J0034--0534 is most ``underluminous'', 
its luminosity, $L_{\rm X}\simeq 0.5\times 10^{30}$ erg s$^{-1}$,
is as low as that estimated for PSR J1024--0719,
the least energetic pulsar among the MSPs discussed in Sections 2--4.
The X-ray efficiency of PSR J0034--0534, 
$\eta_{\rm X}\simeq 1.6\times 10^{-5}$, is smaller at least by an
order of magnitude than those measured for the other MSPs 
(Tables~2 and 3). However, it is worth mentioning
that the estimate on $\ed$
given in Table~2 for PSR J0034--0534 assumes that the intrinsic
period derivative of the pulsar is as directly measured in radio
observations, i.e., without accounting for 
the Shklovskii effect on $\dot{P}$ due to pulsar proper
motion (Shklovskii 1970).
The proper motion of PSR J0034--0534 has not been determined yet. 
Still,
it is reasonable to assume that
the transverse component of the pulsar's velocity
may be about $100\,(d/0.54\,{\rm kpc})^{1/2}$ km s$^{-1}$ 
(what is not unusual for a binary MSP --- see Lommen et al. 2006). 
Then, its intrinsic $\dot{P}$ and,
hence, $\ed$ would be reduced by a factor of 10,
making the pulsar's X-ray efficiency
similar to those found for the MSPs with $\ed<10^{34}$ erg s$^{-1}$.

\section{MSPs and their nebulae}
Relativistic pulsar winds, which carry away pulsar rotational energy, interact
with ambient medium, that is expected to form PWNe
detectable at different wavelengths. In X-rays, 
about 30 PWNe are currently known
(see Kaspi, Roberts \& Harding 2006, Gaensler \& Slane 2006 and
Kargaltsev \& Pavlov 2006 for reviews, as well as electronic PWN catalogs
\footnote{{\tt http://www.astro.psu.edu/users/green/psrdatabase/psrcat.htm},
 {\tt http://www.physics.mcgill.ca/$\sim$pulsar/pwncat.html}}),
thanks mainly to
the superb spatial resolution of \chan\
(see the presentation
by Weisskopf et al. at this conference for a review).
The observed X-ray PWNe are diverse in properties.
Some of them are of a torus-like structure
with jets along the symmetry axis (which apparently coincides
with the pulsar's spin axis,  as suggested for the young 
Crab and Vela pulsars --- Weisskopf et al. 2000, Pavlov et al. 2003).
Several others have a cometary-like shape caused by the pulsar motion.
The ``Mouse'' PWN powered by PSR J1747--2958 (Gaensler et al. 2004)
is confined within a bow-shaped boundary without a shell-like
structure. X-ray PWNe associated with
PSR B1757--24 (Kaspi et al. 2001), J1509 --5859 and J1809--1917
(Sanwal, Pavlov \& Garmire 2005), and the Geminga pulsar
(Caraveo et al. 2003; De Luca et al. 2006; Pavlov, Sanwal \& Zavlin 2006) 
have elongated structures that look like ``trails''
(or ``wakes'') behind the moving pulsars.
These tails could be pulsar jets confined by toroidal magnetic fields,
or they could be associated with shocked relativistic wind
confined by the ram pressure of the surrounding interstellar medium.
Two X-ray nebulae produced by
the MSPs, B1957+20 (Stappers et al. 2003) and J2124--3358 
(Hui \& Becker 2006), belong to the latter group of PWNe.
These are shown in Figures 11 and 12.
Besides, optical observations  
revealed H$_\alpha$ bow-shocks surrounding these two pulsars
(sketched with dashed curves in Figs. 11 and 12).
Both X-ray PWNe 
have a form of a $20''$-long tail, although 
there is a striking difference between their shapes.
The H$_\alpha$ bow-shock and X-ray tail of PSR B1957+20 are
fairly symmetric, with the symmetry axis being nearly aligned with the direction
of the pulsar's proper motion (Fig.~11). 
For PSR J2124--3358, both the bow-shock
and the X-ray tail are bent and highly asymmetric with respect to the 
proper motion vector (Fig.~12). 
Gaensler, John \& Stappers (2002) argued that such an unusual shape of a PWN
could be caused by
a combination of effects associated with anisotropy
of the pulsar wind and density nonuniformity of the surrounding medium.
Analyzing the properties of the X-ray PWN of PSR B1957+20,
Stappers et al. (2003) speculated that the efficiency with which
relativistic particles are accelerated in the pulsar's wind can be
as high as 
those inferred for young pulsars with much stronger surface magnetic
fields. Comparing ratios of X-ray luminosities of tail-like PWNe
to pulsar spin-down energies may be considered as an additional evidence
in favor of this conclusion: the ratio,
$L_{\rm X}/\ed\sim 10^{-4}$,
estimated for the tail associated with PSR B1957+20
is very close to that found for the nebula produced by the 16-kyr-old
PSR B1757--24
(Kaspi et al. 2001)
and greatly exceeds those inferred for the Geminga's tail,
$\sim 4\times 10^{-6}$ (Pavlov, Sanwal \& Zavlin 2006) 
and the Vela's southeast jet,
$\sim 1\times 10^{-6}$ (Pavlov et al. 2003).
Contrary to these two objects,
the third MSP whose supersonic motion through
the ambient medium causes an H$_\alpha$ bow-shock, J0437--4715, 
has not been found to power an X-ray nebula,
suggesting a very low magnetic field in the expected
PWN region (Zavlin et al. 2002).
No an X-ray PWN associated with the most energetic MSPs B1937+21
has been detected in a deep \chan\ observation of this pulsar
(Kuiper et al. 2006).

\begin{figure}
\centering
\includegraphics[height=9.5cm,angle=0]{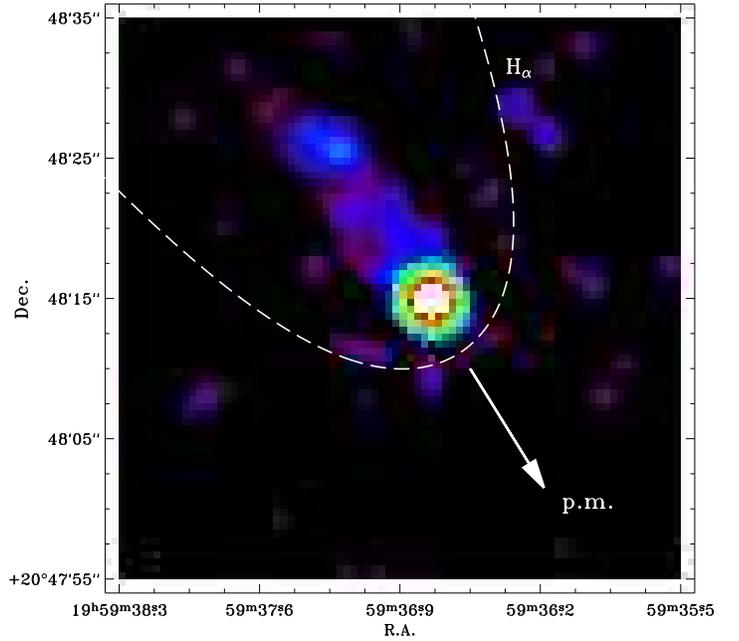}
\caption{\chan\ image of PSR B1957+20 and its tail-like X-ray PWN.
The dashed curve sketches the H$_\alpha$ nebula associated with the pulsar.
The arrow indicates the direction of the pulsar's proper motion.
}
\label{fig:11}
\vskip -0pt
\end{figure}

\begin{figure}
\centering
\includegraphics[height=9.5cm,angle=0]{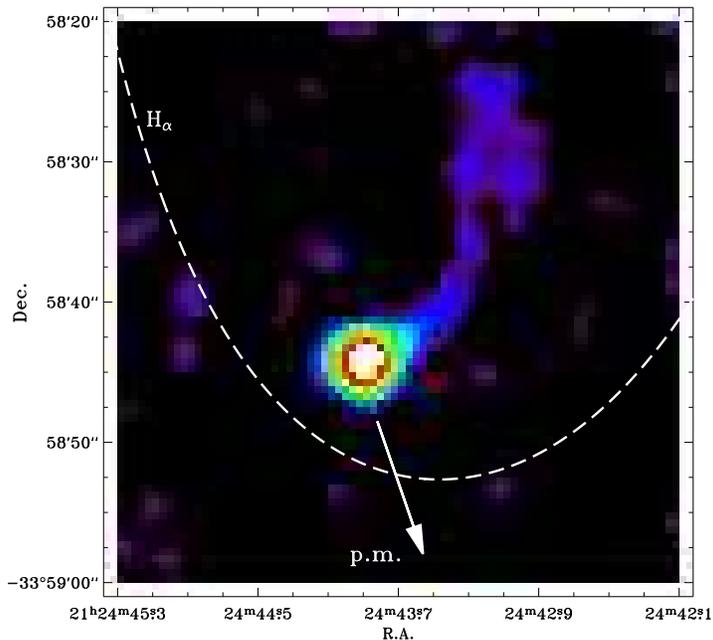}
\caption{Same as in Fig.~11 for PSR J2124--3358.
}
\label{fig:12}
\vskip -0pt
\end{figure}

\section{Summarizing remarks}
X-ray observations of the MSPs discussed in this paper and
those located in the globular clusters
(Bogdanov et al. 2006)
revealed that most of them emit predominantly thermal PC
emission with a luminosity of $\lbolpc\sim 10^{30}$--$10^{31}$ erg s$^{-1}$
and ``PC efficiency'' of $\epc\sim 10^{-4}$--$10^{-3}$.
It is worth noting that these estimates on $\epc$ are close
to that derived for the old ordinary pulsar B0950+08
($\tau\simeq 17$ Myr, $P\simeq 253$ ms,
$\ed\simeq 6\times 10^{32}$ erg s$^{-1}$),
$\epc\simeq 2.4\times 10^{-4}$ (Zavlin \& Pavlov 2004),
suggesting that heating mechanisms in ordinary and rapidly spinning
pulsars may be quite similar.
Comparing pulsar parameters of the MSPs with detected X-ray radiation 
shows that the four MSPs with nonthermal radiation
possess much large values of $\ed$,
by at least a factor of 50, than those estimated for the
thermally emitting pulsars.
The numbers presented in Table~2 indicate that the larger is $\ed$,
the higher is the ``nonthermal'' efficiency $\enon$, whereas the opposite
tendency is apparent for $\epc$.
A similar behavior of $\epc$  follows from results
of the theoretical modeling of the PC heating by returning positrons
produced
through curvature radiation and inverse Compton
scattering (see Figs.~7 and 8 in Harding \& Muslimov
2001 and 2002, respectively, with account for the relation
$\ed\sim\tau^{-[n+1]/[n-1]}=\tau^{-2}$, for the magneto-dipole braking
index $n=3$).
In addition,
high luminosity
of the nonthermal emission may be associated with
large values of the magnetic field at the pulsar light cylinder
(Saito et al. 1997),
$B_{\rm lc}=B_{\rm surf} [2\pi R_{\rm NS}/cP]^3$.
Indeed, the nonthermally emitting
MSPs possess magnetic fields,
$B_{\rm lc}\sim (0.3$--$1)\times 10^6$ G, close to that
of the Crab pulsar ($\sim 1\times 10^6$ G) and
exceeding those in the MSPs
with thermal emission (as well as in PSRs J0751+1807 and J1012+5307),
$B_{\rm lc}\sim 0.03\times 10^6$ G,
at least by an order of magnitude. This may indicate
that the nonthermal radiation of MSPs is generated in
emission zone(s) close to the light cylinder (as predicted
by the outer-gap pulsar models)
and that $B_{\rm lc}$ is an important parameter governing the
magnetospheric activity. On the other hand,
a model by Harding, Usov \& Muslimov (2005)
for acceleration and pair cascades on open field lines
close to pulsar PCs describes rather well
the spectrum of PSR J0218+4232 observed from X-rays through $\gamma$-rays.
Hence, at the present stage,
neither of the two models (outer-gap or polar-cap)
of the nonthermal emission of MSPs
could be regarded as more favorable.
Besides, one cannot rule out that the pulsar
spin period also affects $\enon$ --- the MSPs of the first group have shorter
periods than those of the other four pulsars (Table~1).
Definitely, a bigger sample of X-ray emitting MSPs 
is required to draw more conclusive speculations.

The results of Section~3 point out that PCs of MSPs are likely  
nonuniform, with temperatures decreasing by a factor of 2--3
from the PC center down to its edge. There have not been yet
reliable calculations of the temperature distribution around the pulsar
magnetic poles. 
Hence, sophisticated theoretical
models of PC heating and temperature distribution
are required for further investigation of thermal X-ray
emission from MSPs.

As shown by Pavlov \& Zavlin (1997), modeling of an X-ray pulsed 
profile can yield constraints on
the pulsar mass-to-radius ratio as well as its geometry. 
To do that, more elaborated models
of magnetospheric pulsed emission and data of a better quality at higher
energies are required to disentangle 
nonthermal and thermal components.

\chan\ and \xmm\ provided remarkable detections of
two double-neutron-star binary systems with MSPs as companions,
J0737--3039 and J1537+1155
(Campana, Possentti \& Burgay 2004;
McLaughlin et al. 2004;
Pellizzioni et al. 2004; 
Kargaltsev, Pavlov \& Garmire 2006).
Although the collected data are of a rather scanty statistics,
one could conclude that the inferred X-ray spectra and luminosities
of these two system are similar, suggesting the same mechanisms
generating the detected radiation. 
Among various interpretations proposed,
the most plausible one is that the observed emission consists of
a thermal PC component emitted by the MSPs
plus nonthermal X-rays from the interaction of the pulsar winds
with the neutron star companions.
This interpretation explains a gap observed
in the orbital dependence of the
X-ray flux emitted by the highly eccentric J1537+1155 system 
(Kargaltsev, Pavlov \& Garmire 2006).
To investigate these systems in more detail, a phase-resolved spectroscopy on
X-ray data of a better quality is required to separate possible components
in X-rays from these objects.
Further observations are
expected to constrain both the properties of the pulsar winds
and those of the pulsars themselves in these (and other)
double-neutron-star systems.

So far, only one MSP, J0437--4715,
has been observed and detected in the $UV/FUV$
band (Kargaltsev, Pavlov \& Romani 2004).
The shape of the inferred spectrum suggests thermal emission
from the whole neutron star surface of a surprisingly high
temperature of about 0.1 MK. A heating mechanism should be
operating in a Gyr-old neutron star to keep its surface
at such temperature. 
As discussed by Kargaltsev, Pavlov \& Romani (2004),
among several possible mechanisms (both internal and external), 
chemical heating 
(Reisenegger 1995) and frictional heating (e.g., Cheng et al. 1992)
of the neutron star core are the most plausible options.
To understand thermal evolution of neutron stars, more MSPs
(e.g., the other three pulsars discussed in Sec.~3)
and close ordinary old pulsars (PSR B0950+08 is one of the best
candidates --- Zavlin \& Pavlov 2004) should be observed
in the $UV/FUV$ band.
The example of \psra\ shows that such an observational
program would be feasible.

As a final remark, it is worth mentioning
that so far all known isolated (non-accreting) MSPs were first 
discovered in radio. 
In X-rays a first attempt to find millisecond pulsations from
an isolated compact object was done by Zavlin \& Pavlov (2006)
who suggested that the most famous (and intriguing) member of the
group of the ``dim'' isolated, radio-quite, and thermally emitting
neutron stars discovered with $ROSAT$ (see
the presentation by F. Haberl at this conference), RX~J1856.5--3754, 
could be an MSP (see Pavlov \& Zavlin 2003
for a discussion). One of the reasons to look for
millisecond pulsations was the fact that the
X-ray flux detected from this object in deep observations with \chan\
and \xmm\ revealed no pulsations at periods above 20 ms (the period range
to which those observational data were sensitive), with a stringent upper
limit on the pulsed fraction $\fp<1.3\%$.
To check the MSP hypothesis, a deep \xmm\ observation of 
RX~J1856.5--3754 in an instrumental model with a temporal
resolution of 0.03 ms 
was conducted in April 2004. Despite a very large number of source
counts collected in this observation ($\sim 2\times 10^5$),
no significant pulsations were found at periods in the range $P=1$--20 ms.
The derived 1$\sigma$ upper limit on the pulsed flux for this period
range is $\fp<2.1\%$. This nondetection virtually excludes the MSP
interpretation for RX~J1856.5--3754 and supports alternative hypotheses
of a very special neutron star's geometry and/or a very long spin period
($P\gapr 10$ hr).

\begin{acknowledgements}
\vskip -10pt
The author thanks George Pavlov for useful discussions.
This work is supported by a NASA Research Associateship Award 
at NASA Marshall Space Flight Center.
The study of RX~J1856.5--3754 with \xmm\ was made possible through
the NASA grant NNG04GI801G.
\vskip -0pt
\end{acknowledgements}



\end{document}